# Critical components in 0.14 THz communication systems

Guangcun Shan[1,*], Xinghai Zhao[2], Haoshen Zhu[3], and Chan-Hung Shek[1,*]

1 Department of Physics and Materials Sciences, City University of Hong Kong, Hong Kong SAR
Email: guangcunshan@mail.sim.ac.cn,
2 Institute of Electronic Engineering, China Academy of Engineering Physics, Sichuan, China
3 Department of Electronic Engineering, City University of Hong Kong, Hong Kong SAR

*Abstract* — In the super-heterodyne terahertz communication system, the proper design of the critical components like mixers and filters are of great importance for enhancing its performance. In this work, some issues on our newly developed system setup design for 0.14 THz wireless communications and the key components subharmonic mixer (SHM) based on Schottky diode, as well as silicon micromachined bandpass rectangular waveguide filters are presented. According to ADS simulation, the optimum conversion loss of the 140 GHz SHM is around 8.2 dB. And the silicon-micromachined rectangular waveguide filters have been fabricated and the measured lowset insertion losses are lower than 0.5 dB.

## I. INTRODUCTION

Terahertz (THz) science and technology in the working frequency of 0.1 – 10 THz has become a R&D hotspot in recent years due to both its scientific interest and potential applications in many areas such as genetic research, biochemical material identification, medical imaging, and security screening, etc[1-3]. THz communication refers to the use of THz-band wireless communication (0.1 THz~10 THz) as a carrier for space communications. Compared to conventional microwave optical communications, THz communications has advantages of broadband, good directionality, strong anti-interference ability, privileged communication confidentiality and some other characteristics. Moreover, THz communications with high energy efficiency, penetrating power are ideal for broadband short-distance wireless transmission, high-speed wireless transmission and space satellite communications, and some other fields.

Typical existing THz communications research results are: Brauns-chweig THz communications laboratory in Germany has developed an experiment system based on room temperature two dimensional electron gas (2DEG) modulator and femtosecond laser THz time-domain spectroscopy instrument for communication [3], and Schottky harmonic mixing techniques based 0.3 THz wireless communication system[4]; Japan NTT has developed the optical UTC-PD [5] based and solid-state monolithic microwave integrated circuit (MMIC)-based 120 GHz 10 Gbps wireless communication systems[6]. Moreover, researchers from China Academy of Engineering Physics and City University of Hong Kong have proposed 0.14 THz wireless communication system by using subterahertz transmitter and receiver and 140GHz silicon micromachined bandpass rectangular waveguide filters [7,8]. The main difficulties encountered for current main THz communications are difficulty of modulation and demodulation techniques, and low radiated power, as well as severe atmospheric extinction. Besides, the required compact components for such systems like planar integrated sources, amplifiers and antenna arrays do not exist yet. In particular, we have recently proposed and implemented one novel 0.14THz wireless communications system, which is based on the superheterodyne system with high spectral efficiency, means of resistance to distortion.

In this work, we would present our newly developed system setup design for 0.14 THz wireless communications and the key components for 140 GHz wireless transmitter and receiver, e.g., Schottky diode-based sub-harmonic mixers (SHM) using Quartz substrate microstrip junction structure with ADS simulation method, test results show that the single side band (SSB) conversion loss 8.2 dB as well as the fabrication and performance of micro electro mechanical systems (MEMS) technology based 0.14 THz rectangular band-pass waveguide filter.

## II. SYSTEM AND COMPONENTS

The proposed transmission system consists of autarkic transmitter and receiver units as shown in the block diagram in Fig. 1. A SHM is used to upconvert a signal (0 - 10 GHz, delivered by a signal generator) up to 140 GHz. It is then transmitted with a horn antenna directly attached to the mixer block (with the output power of 50 mW for the 140 GHz carrier signal). The local oscillation signal is provided by a 32.6 GHz phase-locked dielectric resonator oscillator (DPRO with 10 MHz reference crystal oscillator), which is first amplified and doubled. At the receiver side the same components have been used except that the DPRO is tuned to 32 GHz. This results in a downconversion of the received signal to an intermediate frequency (IF) of 2.4 GHz. The upper sideband of the downconverted signal is displayed on a spectrum analyzer in the frequency range 2.4–10 GHz.

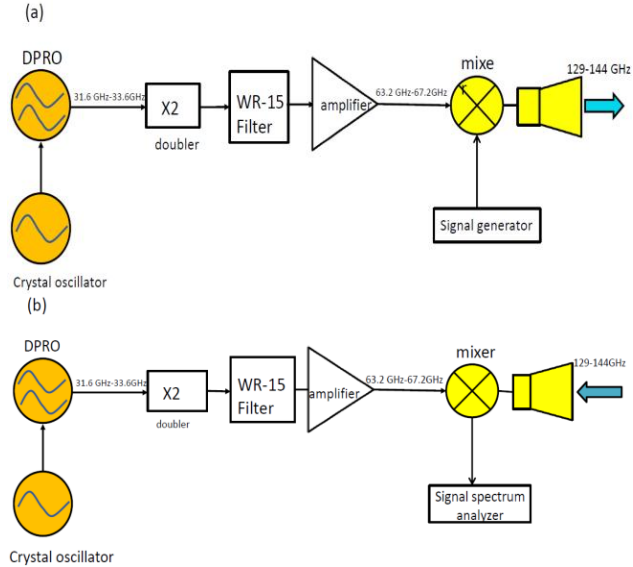

Fig. 1. Block diagram for transmitter and receiver units of 140 GHz communication system.

The millimeterwave mixer is the key component to realize frequency conversion in THz communication systems. Therefore, it has been extensively explored in various research institutes including United Kingdom RAL Laboratories, NASA JPL Laboratory, United States University of Virginia, VIRGINIA Diodes Inc. (VDI) etc. Thomas et al. have developed 330 GHz harmonic mixer [6] to achieve 5.7 dB DSB (Double Side Band, DSB) conversion loss, noise temperature is less than 930 K; Jeffrey L Hesler et al. have developed 585 GHz harmonic mixer [7], DSB conversion loss reached 7.3 dB noise temperature 2380 K; Mehdi I Human design 640 GHz harmonic mixers, enabling 8.1 dB DSB Conversion loss and 1640 K DSB Noise temperature [8]. In a SHM the mixing mechanism is conducted between the radio frequency (RF) or IF signals and one of the harmonics of the local oscillator (LO). Thus, the nonlinear device (diode, metal semiconductor field effect transistor, etc.) performs both mixing and frequency multiplication. The SHMs are very sensitive to the loading of the nonlinear device at the various idler frequencies. Compared to the SIS HEB mixer, Schottky diode mixer outperforms with some distinctive features such as working at room temperature, low cost, low conversion loss, moderate noise temperature, and etc. These mixers are very economical at millimeter wave frequencies since low frequency low cost microwave sources (as LO) can be adopted as higher-order harmonics will perform Therefore, they are of great interest for the current millimeter-wave THz frequency band transmitter and receiver. Various SHMs working at millimeter wave frequencies have been successfully demonstrated by some researches [9]-[11].

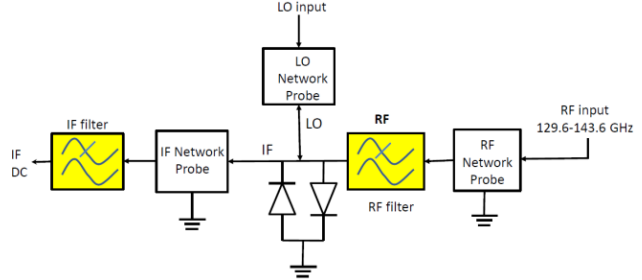

Fig. 2. Schematic structure of 140GHz sub-harmonic mixer.

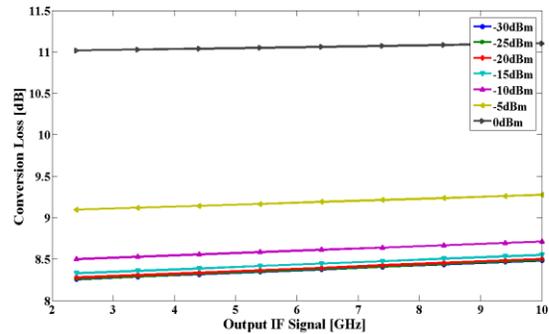

Fig. 3. Simulated result of 140GHz sub-harmonic mixer.

A 140 GHz SHM is designed based on the anti-parallel UMS DBES105a Schottky diode pair. This design shows the advantageous suppression of odd harmonics during mixing. In order to simultaneously achieve low conversion loss and good matching, the structure shown in Fig. 2 is implemented. The mixer model with idealized matching networks is simulated in ADS to verify the low conversion loss second-harmonic mixing concept. The mixer model with idealized matching networks is simulated in ADS to verify the low conversion loss second-harmonic mixing concept. With a fixed LO input power (6dBm), this SHM achieves an optimum conversion loss at around 8.25dB. As illustrated in Fig. 3, the conversion loss is slightly changed as output IF signal in the receiver unit varying from 2.4GHz to 10GHz for input RF carrier with various frequencies but constant power. Moreover, according to simulation result, it is shown that with an increased input RF power and fixed LO power, the conversion loss is enlarged dramatically (around 11dB with 0dBm RF signal). This indicates that other mixing products become significant and the ideal received RF power should be kept around -20dBm.

In the process of frequency conversion in harmonic mixers, upper and lower sideband, development needs 140 GHz BPF to filter out lower side band signal to avoid image interference receiver. Rectangular waveguides are typical transmission lines used in microwave and millimeter-wave applications. The advantages of the rectangular waveguides include wide bandwidths of operation for single-mode transmission, low attenuation, and good mode stability for the fundamental mode of propagation. In terms of RF performance, the metal-pipe rectangular waveguide can be considered as the ultimate guided-wave structure, although package density is very poor indeed. The micromachining techniques for very high frequency or sub-THz devices relates to deep reactive ion etching (DRIE), LIGA, laser micromachining and SU-8 technology[7,14-19].

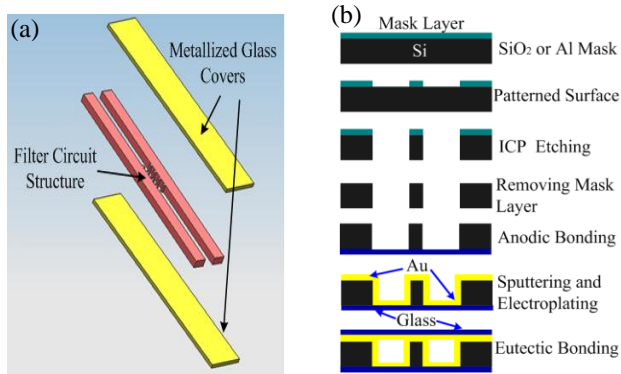

Fig. 4. (a) The schematic of filter assembly. The metallized glass covers are bonded on top and bottom of the open waveguide to form closed cavity. (b) Main microfabrication assembly flow.

Recently, we have fabricated cutting-edge silicon micromachined D-band 3-pole and 5-pole bandpass filters using inductively-coupled plasma (ICP) dry etching technology[7,8]. The initial geometrical structure parameters were calculated by the 0.5dB ripple 3-pole and 5-pole Chebyscheff filter model. The rectangular waveguide in the Ansoft's HFSS simulation model has a cross-section of 1.651mm×0.825 mm and the thickness of iris is 0.1mm. And the simulation was set up using gold layers for all of the internal walls of the waveguide. All attempts in our simulation were made to reasonably account for thickness and roughness in the simulations due to the etching and gold depositing process in order to obtain much more accurate results. The simulated insertion losses in the simulation are ~0.25dB for 3-pole filter and ~0.35dB for 5-pole filter, and the filter bandwidth is approximate to be 21GHz (15%) and 16GHz (11.5%), respectively. The isolation is larger than 18dB.

To fabricate an iris waveguide filter, one solution is to fabricate two identical parts and then to bond them together [16]. Obviously, this method requires a precise alignment at bonding. Note that complex process steps leading to low device reliability and augmenting the device cost would hinder the application of micromachining technologies. Hence, E-plane irises are chosen to form the iris waveguide filter because an appropriate aspect ratio and a process without aligned-bonding as shown in Fig.4 (a) could be implemented. The 140GHz MEMS filter irises were fabricated by the ICP reactive ion etcher (AMS100SE, Alcatel Co.) and then the filter circuit structure were bonded together with the metallized glass covers to form the waveguide cavity [Fig. 4(a)]. The microfabrication process is shown in Fig.4 (b). The waveguide filter was fabricated on silicon wafer with a thickness of 825 μm. The iris pattern was then etched through the full thickness of the thick polished silicon wafer. The through-wafer etching procedure is accomplished using a deep reactive ion etcher with a repeating etch ($SF_6$) and passivation ($C_4F_8$) cycle, commonly referred to as the Bosch process [7]. And bonding method to add a rectangular waveguide cover up and down, the last forming filter in dicing and assembly to the fixture device.

Fig.5 (a) and (b) shows the measured and simulated responses of the prototype 3-pole filter and those of the 5-pole filter between 125 GHz and 170 GHz, respectively. As shown in Fig. 5 (a), the 3-pole filter shows a bandwidth of 21.2 GHz centered at 141.2 GHz, while our simulation results predicts a bandwidth of 21 GHz centered at 140 GHz. In contrast, as shown in Fig. 5 (b), the 5-pole prototype filter has the insertion loss of 0.61 dB with a real bandwidth 14.5 GHz, and a further estimated unloaded quality factor for a single cavity resonator is 815. Our experimental results show that it is a potential reliable solution for future subterahertz waveguide RF devices volume fabrication. We have demonstrated our approach using ICP dry etching technology to fabricate a 140GHz waveguide filter. This technology is good at etching deep, anisotropic structures in silicon. The measured insertion loss is about 0.4~0.7dB achieved by the frequency multiplication way and the measured bandwidth error is less than 5% with the central frequency of 140GHz ±2GHz, which is in accord with the design and simulation results. The etched trench deep can get to be larger than 800μm with a sidewall slope of ≥89% and the surface roughness can be improved to be less than 200 nm by optimizing ICP etching and electroplating processes. The fabrication inaccuracy is less than ±0.03mm in the filter dimensions, besides that the iris thickness varies by less than 10%. These imply that the fabrication technique and method using ICP dry etching technology for one-step forming and volume production has the potential for operation in the sub-millimeter and sub-THz regime with precisely designed RF device elements.

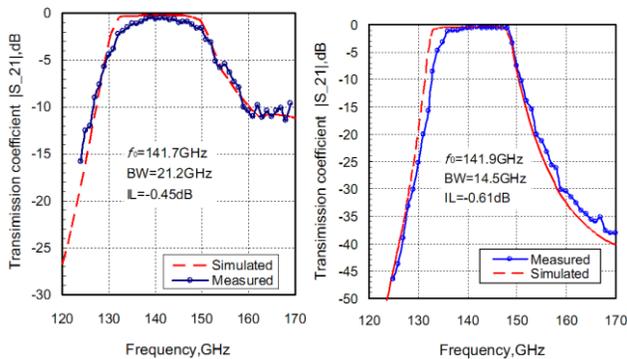

Fig. 5.Measured and simulated results of (a) 3-pole sample with iris thickness of 105μm and (b) 5-pole sample with iris thickness of 110μm

III. CONCLUSION

This article describes the recent development of the system setup and the key components in 140 GHz wireless commnucations. We have explored the Schottky diode based SHM, on the basis of HFSS+ADS joint simulation and filter matrix integration design method of launched a quartz substrate by sputter process, and 140GHz rectangular waveguides using high accuracy micromachining MEMS technology. The 0.14THz wireless system study is still in its infancy, the development will be continued to improve in subsequent studies.

ACKNOWLEDGEMENT

The authors wish to acknowledge Dr. J.F. Bao for helpful measurement on rectangular waveguide filter.